\documentstyle[preprint, aps]{revtex}
\begin{document}
\title{Interface dynamics equations: thier properties and computer simulation.}
\author{V.Gafiychuk, A.Shnyr, B.Dacko}
\author{Institute of Applied Problems of Mechanics and Mathematics,}
\author{National Academy of Science of Ukraine}
\date{\today }
\maketitle

\begin{abstract}
We investigate the interface dynamic in Laplacian growth model, using the
conformal mapping technique. Starting from the governig equation for the
conformal map, obtained by B.Shraiman and D.Bensimon, we derive different
possible forms of the equation. Some of them have evident physical
representation and are convenient for computer simulation, the results of
which are presented in this paper.
\end{abstract}

Formation and evolution of interfaces in systems with phase transition of
the first kind led to one of the most striking phenomena of
self-organization -- the growth of dendritic structures. These phenomena are
inherent to many branches of physics, chemistry and biology and are quite
similar from mathematical point of view. They are known as the moving
boundary problems or Stefan problems \cite{free,free1}.

Such problems, as usual, have no analytical solutions caused by the
essential non-linearity stipulated by the representation of the boundary
conditions for parameters, allocated in space on the free boundary. In the
given class of systems with free boundaries it is possible to single out a
subclass where the distribution of the field in the space is stationary. It
is known as quasistationary Stefan problem.\ Usage of the methods of
conformal mapping allows to reduce the last problem to a dynamics of some
physical fields in the system with fixed boundaries. This approach is more
suitable for analytical and computer simulation.

Theoretical methods of the functions with the complex variable appear to be
rather useful in the construction of the effective surface dynamics of the
interface at least in the two-dimensional problem. In this case the central
part belongs to the conformal mappings of the physical regions of the space
onto some special kind of regions. A large number of models for interfacial
dynamics have been proposed by now. Reviews of these problems can be found
in \cite{pec,ben1,glb}. The most famous equation has been introduced by
Shraiman and Bensimon \cite{sb} and now is widely known as the conformal
mapping equation and it is still under investigation.

Let us consider the problem of Laplacian growth, which can be formulated in
the following way. We investigate the growth of two-dimensional region $D$
and suppose, that its boundary $\partial D$, which represents the physical
interface, is the analytic Jordan curve $\Gamma $ \ref{fig1}. The field $%
\varphi $ outside of the region $D$, obeys the Laplace equation 
\begin{equation}
\Delta \varphi =0.  \label{f1}
\end{equation}
The boundary $\partial D$ grows at a rate, that is proportional to the
normal gradient of the field of the interface. Therefore the evolution of
the interface is governed by the following equation 
\begin{equation}
\upsilon _{n}=\left. -\nabla _{{\bf n}}\varphi \right| _{\Gamma }.
\label{f2}
\end{equation}
The potential field $\varphi $ satisfies the following boundary condition at
the interface 
\begin{equation}
\left. \varphi \right| _{\Gamma }=\left. d_{0}k\right| _{\Gamma },
\label{f3}
\end{equation}
where $k$ is the curvature of the interface $\Gamma $, $d_{0}$ -- is the
dimensionless surface tension parameter.

\begin{figure}[tbp]
\caption{Time-dependent conformal mapping from the exterior of the unit disc
onto the domain of our interest.}
\label{fig1}
\end{figure}

Equations (\ref{f1})-(\ref{f3}) determine the boundary-value problem with
moving interface, or Stefan problem. Usage of the  the methods of conformal
mapping \cite{ben1,glb,sb,blu,pet,men,ben,gl1,gl2,derH}, which are based on
the Riemann mapping theorem, allows to simplify the considered task. \ The
theorem states the existence of a conformal map from the exterior of $D$
onto a standard domain, for example, the exterior of the unit disk, and
boundary $\partial D$ corresponds to unit circle. In outline one can
parametrize the evolution of the interface in $z$ plane by the
time-dependent conformal map $F\left( w,t\right) $, which takes the exterior
of the unit disk at each instant of time, $\left| w\right| =1$, onto the
exterior of $D$. Therefore the evolution of the interface can be represented
in the terms of 
\begin{equation}
\Gamma \left( \theta ,t\right) =\lim_{w\rightarrow e^{i\theta }}F\left(
w,t\right) ,\text{ }0\leq \theta <2\pi   \label{f4}
\end{equation}
As it is shown by Shraiman and Bensimon \cite{sb}, in case of free surface
tension $\left( d_{0}=0\right) $ the conformal map fulfills the following
equation of motion 
\begin{equation}
\frac{\partial \Gamma \left( \theta ,t\right) }{\partial t}=-i\frac{\partial
\Gamma \left( \theta ,t\right) }{\partial \theta }\left. S\left[ \left| 
\frac{\partial \Gamma \left( \theta ,t\right) }{\partial \theta }\right|
^{-2}\right] \right| _{w=e^{i\theta }},  \label{f5}
\end{equation}
where $S\left[ \cdot \right] $ stands for Schwartz operator. For the
exterior of the unit circle with some boundary condition $f\left( \theta
\right) $\ the Schwartz operator can be represented in the following way 
\begin{equation}
S\left[ f\left( \theta \right) \right] =-\frac{1}{2\pi }\int\limits_{0}^{2%
\pi }d\theta ^{\prime }\frac{e^{i\theta ^{\prime }}+w}{e^{i\theta ^{\prime
}}-w}f\left( \theta ^{\prime }\right) +iC,  \label{sw}
\end{equation}
where $C$ is an arbitrary constant. Denoting the action of the Schwartz
operator on $\left| \frac{\partial \Gamma \left( \theta ,t\right) }{\partial
\theta }\right| ^{-2}$ as the limit of an analytic function $G\left(
w,t\right) $, which explicit form is unique and explicitly determinable \cite
{blu}, the equation of motion for $F\left( w,t\right) $ (\ref{f5}) is
transformed to 
\begin{equation}
\frac{dF\left( w,t\right) }{dt}=w\frac{dF\left( w,t\right) }{dw}G\left(
w,t\right) .  \label{f6}
\end{equation}
Let us also write down the logarithmic derivative of the (\ref{f6}) with
respect to $w$%
\begin{equation}
\frac{d}{dt}\ln \frac{dF\left( w,t\right) }{dw}=\left( \frac{dF\left(
w,t\right) }{dw}\right) ^{-1}\frac{d}{dw}\left( w\frac{dF\left( w,t\right) }{%
dw}G\left( w,t\right) \right) .  \label{f7}
\end{equation}
Using the explicit forms of the conformal map $F\left( w,t\right) $ one can
reduce the partial differential equation (\ref{f5}), governing the interface
dynamics, to a set of ordinary differential equations for the dynamics of
the critical points of the conformal map \cite{sb}. This facilitates the
analytical and numerical investigations of the complex behavior in the
evolution of the interface \cite{blu,pet,men}.

In spite of this fact sometimes it is convenient to introduce the following
field \cite{glb,gl1,gl2} for these investigations 
\begin{equation}
\eta \left( \theta ,t\right) =\left| \frac{dF\left( w,t\right) }{dw}\right|
_{w=e^{i\theta }}^{-1},  \label{f8}
\end{equation}
which has clear representation -- the stretch of the interface $\Gamma
\left( \theta ,t\right) $ under conformal map $F^{-1}\left( z,t\right) $.
The conformality of $F\left( w,t\right) $ demands that the critical points
of the map, zeroes of its derivative, all lie within the unit disk.
Therefore $\ln \frac{dF\left( w,t\right) }{dw}$ is an analytic function
outside of the region $D$. Regarding that 
\begin{equation}
\ln \left. \left[ \frac{dF\left( w,t\right) }{dw}\right] \right|
_{w=e^{i\theta }}=\ln \left. \left| \frac{dF\left( w,t\right) }{dw}\right|
\right| _{w=e^{i\theta }}+i\left. \psi \left( w,t\right) \right|
_{w=e^{i\theta }},  \label{f9}
\end{equation}
and solving Schwartz problem outside the unit disc in $w$ plane, one can
recover the equation of motion for $F\left( w,t\right) $ from the equation
of motion for $\eta \left( \theta ,t\right) $, as well as from $\psi \left(
\theta ,t\right) $ one ($\psi \left( \theta ,t\right) $ represents oriented
angle between tangent vectors $\tau _{w}$ and $\tau _{F\left( w\right) }$ 
\ref{fig1}). To keep clear the sense of using (\ref{sw}) we should also
write down the relations between $\eta \left( \theta ,t\right) $ and $\psi
\left( \theta ,t\right) $ \cite{glb,gl2} 
\begin{equation}
\psi \left( \theta ,t\right) =-\frac{1}{2\pi }\int\limits_{0}^{2\pi }d\theta
^{\prime }\coth \frac{\theta ^{\prime }-\theta }{2}\ln \eta \left( \theta
^{\prime },t\right) .  \label{f10}
\end{equation}

In the present paper we investigate the properties of the equation of
motion, obtained by Shraiman and Bensimon (\ref{f5}), and derive the
equation of motion for $\eta \left( \theta ,t\right) $, that is more
suitable for numerical simulation. In this case let us note that 
\begin{equation}
\left| \frac{\partial \Gamma \left( \theta ,t\right) }{\partial \theta }%
\right| ^{-2}=\left| \frac{dF\left( w,t\right) }{dw}\frac{dw}{d\theta }%
\right| _{w=e^{i\theta }}^{-2}=\left| \frac{dF\left( w,t\right) }{dw}\right|
_{w=e^{i\theta }}^{-2}=\eta ^{2}\left( \theta ,t\right) ,  \label{f11}
\end{equation}
thus $G\left( w,t\right) =S\left( \eta ^{2}\left( \theta ,t\right) \right) $%
, where all constants in Schwartz operator are explicitly determinable \cite
{glb,blu}. Regarding the real part of the left side of the\ equation (\ref
{f7}) as $\frac{d}{dt}\ln \left| \frac{dF\left( w,t\right) }{dw}\right| $
and simplifying the right side of the equation one can get 
\begin{equation}
\frac{d}{dt}\ln \left| \frac{dF\left( w,t\right) }{dw}\right| =%
%TCIMACRO{\func{Re}}%
%BeginExpansion
\mathop{\rm Re}%
%EndExpansion
\left[ G\left( w,t\right) +w\frac{dG\left( w,t\right) }{dw}+wG\left(
w,t\right) \frac{d}{dw}\ln \frac{dF\left( w,t\right) }{dw}\right] .
\label{f13}
\end{equation}
Determining the real part for each term within square brackets of the right
side of (\ref{f13}) for $w=e^{i\theta }$%
\begin{equation}
%TCIMACRO{\func{Re}}%
%BeginExpansion
\mathop{\rm Re}%
%EndExpansion
\left. \left[ G\left( w,t\right) \right] \right| _{w=e^{i\theta }}=\eta
^{2}\left( \theta ,t\right) ,  \label{f14}
\end{equation}
\begin{equation}
%TCIMACRO{\func{Re}}%
%BeginExpansion
\mathop{\rm Re}%
%EndExpansion
\left. \left[ w\frac{dG\left( w,t\right) }{dw}\right] \right| _{w=e^{i\theta
}}=\frac{\partial }{\partial \theta }%
%TCIMACRO{\func{Im}}%
%BeginExpansion
\mathop{\rm Im}%
%EndExpansion
\left. \left[ G\left( w,t\right) \right] \right| _{w=e^{i\theta }},
\label{f15}
\end{equation}
\begin{eqnarray}
&&%
%TCIMACRO{\func{Re}}%
%BeginExpansion
\mathop{\rm Re}%
%EndExpansion
\left. \left[ wG\left( w,t\right) \frac{d}{dw}\ln \frac{dF\left( w,t\right) 
}{dw}\right] \right| _{w=e^{i\theta }}  \nonumber \\
&=&%
%TCIMACRO{\func{Re}}%
%BeginExpansion
\mathop{\rm Re}%
%EndExpansion
\left. \left[ wG\left( w,t\right) \frac{d}{dw}\left( \ln \left| \frac{%
dF\left( w,t\right) }{dw}\right| +i\psi \left( w,t\right) \right) \right]
\right| _{w=e^{i\theta }}=  \nonumber \\
&=&-\eta ^{2}\left( \theta ,t\right) \frac{\partial \psi \left( \theta
,t\right) }{\partial \theta }+%
%TCIMACRO{\func{Im}}%
%BeginExpansion
\mathop{\rm Im}%
%EndExpansion
\left[ G\left( w,t\right) \right] \frac{d}{d\theta }\ln \left. \left| \frac{%
dF\left( w,t\right) }{dw}\right| \right| _{w=e^{i\theta }},  \label{f16}
\end{eqnarray}
and taking into account (\ref{f8}), (\ref{f11}) we rewrite the equation (\ref
{f13}) in the following form 
\begin{eqnarray}
\frac{d}{dt}\ln \eta ^{-1}\left( \theta ,t\right) &=&\eta ^{2}\left( \theta
,t\right) +\frac{\partial }{\partial \theta }%
%TCIMACRO{\func{Im}}%
%BeginExpansion
\mathop{\rm Im}%
%EndExpansion
\left. \left[ G\left( w,t\right) \right] \right| _{w=e^{i\theta }}-\eta
^{2}\left( \theta ,t\right) \frac{\partial \psi \left( \theta ,t\right) }{%
\partial \theta }  \nonumber \\
&&+%
%TCIMACRO{\func{Im}}%
%BeginExpansion
\mathop{\rm Im}%
%EndExpansion
\left. \left[ G\left( w,t\right) \right] \right| _{w=e^{i\theta }}\frac{d}{%
d\theta }\ln \eta ^{-1}\left( \theta ,t\right) .  \label{f17}
\end{eqnarray}
Taking into account (\ref{f10}) and the fact, that imaginary part of
Schwartz problem solution (\ref{sw}) can be written in the terms of 
\begin{equation}
%TCIMACRO{\func{Im}}%
%BeginExpansion
\mathop{\rm Im}%
%EndExpansion
\left. \left[ G\left( w,t\right) \right] \right| _{w=e^{i\theta }}=\frac{1}{%
2\pi }\int\limits_{0}^{2\pi }d\theta ^{\prime }\coth \frac{\theta ^{\prime
}-\theta }{2}\eta ^{2}\left( \theta ^{\prime },t\right) ,  \label{f18}
\end{equation}
one can simplify equation (\ref{f17}) and obtain the following equation of
motion for $\eta \left( \theta ,t\right) $%
\begin{eqnarray}
\frac{\partial \eta \left( \theta ,t\right) }{\partial t} &=&-\eta
^{3}\left( \theta ,t\right) -\eta \left( \theta ,t\right) \frac{\partial }{%
\partial \theta }\frac{1}{2\pi }\int\limits_{0}^{2\pi }d\theta ^{\prime
}\coth \frac{\theta ^{\prime }-\theta }{2}\eta ^{2}\left( \theta ^{\prime
},t\right) +  \nonumber \\
&&+\eta ^{3}\left( \theta ,t\right) \frac{\partial }{\partial \theta }\frac{1%
}{2\pi }\int\limits_{0}^{2\pi }d\theta ^{\prime }\coth \frac{\theta ^{\prime
}-\theta }{2}\ln \eta \left( \theta ^{\prime },t\right) +  \nonumber \\
&&+\frac{\partial \eta \left( \theta ,t\right) }{\partial \theta }\frac{1}{%
2\pi }\int\limits_{0}^{2\pi }d\theta ^{\prime }\coth \frac{\theta ^{\prime
}-\theta }{2}\eta ^{2}\left( \theta ^{\prime },t\right) .  \label{f19}
\end{eqnarray}

The last equation has similar form to the derived one, using the variational
principles of conformal maps \cite{glb,gl1,gl2}.

In order to derive the proper equation of motion for $\eta \left( \theta
,t\right) $ in the presence of surface tension $\left( d_{0}\neq 0\right) $
we should notice, that all we need is to obtain the proper form of $G\left(
w,t\right) $ for the nonzero surface tension\cite{ben}.

Thus, one can rewrite the desired evolution equation for mapping (\ref{f6})
in the following form: 
\begin{equation}
\frac{dF\left( w,t\right) }{dt}=-w\frac{dF\left( w,t\right) }{dw}G_{0}\left(
w,t\right) ,  \label{f23}
\end{equation}
where $G_{0}\left( w,t\right) $ is an analytic function of $w$, which real
part on $\left| w\right| =1$ is specified as follows \cite{ben}: 
\begin{equation}
%TCIMACRO{\func{Re}}%
%BeginExpansion
\mathop{\rm Re}%
%EndExpansion
\left[ G_{0}\left( w,t\right) \right] |_{w=e^{i\theta }}=\left[ \frac{1-d_{0}%
%TCIMACRO{\func{Re}}%
%BeginExpansion
\mathop{\rm Re}%
%EndExpansion
w\frac{dS\left[ k\left( w,t\right) \right] }{dw}}{\left| w\frac{dF\left(
w,t\right) }{dw}\right| ^{2}}\right] _{w=e^{i\theta }}.  \label{f24}
\end{equation}
Comparing equations (\ref{f23}) and (\ref{f6}) we should notice, that to
derive the equation of motion for the properly introduced $\eta \left(
\theta ,t\right) ,$ we have to define the real and imaginary parts of $%
\left. G_{0}\left( w,t\right) \right| _{w=e^{i\theta }}$%
\begin{equation}
%TCIMACRO{\func{Re}}%
%BeginExpansion
\mathop{\rm Re}%
%EndExpansion
\left. \left[ G_{0}\left( w,t\right) \right] \right| _{w=e^{i\theta }}=\eta
^{2}\left( \theta ,t\right) \left( 1-d_{0}\frac{\partial }{\partial \theta }%
%TCIMACRO{\func{Im}}%
%BeginExpansion
\mathop{\rm Im}%
%EndExpansion
\left. \left[ S\left[ k\left( w,t\right) \right] \right] \right|
_{w=e^{i\theta }}\right)  \label{f25}
\end{equation}
\begin{equation}
%TCIMACRO{\func{Im}}%
%BeginExpansion
\mathop{\rm Im}%
%EndExpansion
\left. \left[ G_{0}\left( w,t\right) \right] \right| _{w=e^{i\theta }}=\frac{%
1}{2\pi }\int\limits_{0}^{2\pi }d\theta ^{\prime }\coth \frac{\theta
^{\prime }-\theta }{2}\eta ^{2}\left( \theta ^{\prime },t\right) \left(
1-d_{0}\frac{\partial }{\partial \theta ^{\prime }}%
%TCIMACRO{\func{Im}}%
%BeginExpansion
\mathop{\rm Im}%
%EndExpansion
\left. \left[ S\left[ k\left( w,t\right) \right] \right] \right|
_{w=e^{i\theta ^{\prime }}}\right) ,  \label{f26}
\end{equation}
where 
\begin{equation}
%TCIMACRO{\func{Im}}%
%BeginExpansion
\mathop{\rm Im}%
%EndExpansion
\left. \left[ S\left[ k\left( w,t\right) \right] \right] \right|
_{w=e^{i\theta }}=\frac{1}{2\pi }\int\limits_{0}^{2\pi }d\theta ^{\prime
}\coth \frac{\theta ^{\prime }-\theta }{2}k\left( \theta ^{\prime },t\right)
.  \label{f27}
\end{equation}
After the similar simplifications for the free-tension case one could obtain
the following equation of motion for $\eta \left( \theta ,t\right) $: 
\begin{eqnarray}
\frac{\partial \eta \left( \theta ,t\right) }{\partial t} &=&-\eta
^{3}\left( \theta ,t\right) \left( 1-d_{0}\frac{\partial }{\partial \theta }%
\frac{1}{2\pi }\int\limits_{0}^{2\pi }d\theta ^{\prime }\coth \frac{\theta
^{\prime }-\theta }{2}k\left( \theta ^{\prime },t\right) \right) +  \nonumber
\\
&&-\eta \left( \theta ,t\right) \frac{\partial }{\partial \theta }\frac{1}{%
2\pi }\int\limits_{0}^{2\pi }d\theta ^{\prime }\coth \frac{\theta ^{\prime
}-\theta }{2}\eta ^{2}\left( \theta ^{\prime },t\right) \left( 1-d_{0}\frac{%
\partial }{\partial \theta ^{\prime }}\frac{1}{2\pi }\int\limits_{0}^{2\pi
}d\theta ^{\prime \prime }\coth \frac{\theta ^{\prime \prime }-\theta
^{\prime }}{2}k\left( \theta ^{\prime \prime },t\right) \right) +  \nonumber
\\
&&+\eta ^{3}\left( \theta ,t\right) \left( 1-d_{0}\frac{\partial }{\partial
\theta }\frac{1}{2\pi }\int\limits_{0}^{2\pi }d\theta ^{\prime }\coth \frac{%
\theta ^{\prime }-\theta }{2}k\left( \theta ^{\prime },t\right) \right) 
\frac{\partial }{\partial \theta }\frac{1}{2\pi }\int\limits_{0}^{2\pi
}d\theta ^{\prime }\coth \frac{\theta ^{\prime }-\theta }{2}\ln \eta \left(
\theta ^{\prime },t\right) +  \nonumber \\
&&+\frac{\partial \eta \left( \theta ,t\right) }{\partial \theta }\frac{1}{%
2\pi }\int\limits_{0}^{2\pi }d\theta ^{\prime }\coth \frac{\theta ^{\prime
}-\theta }{2}\eta ^{2}\left( \theta ^{\prime },t\right) \left( 1-d_{0}\frac{%
\partial }{\partial \theta ^{\prime }}\frac{1}{2\pi }\int\limits_{0}^{2\pi
}d\theta ^{\prime \prime }\coth \frac{\theta ^{\prime \prime }-\theta }{2}%
k\left( \theta ^{\prime \prime },t\right) \right) ,  \label{f28}
\end{eqnarray}
where $k\left( \theta ,t\right) $ is the curvature of $\Gamma \left( \theta
,t\right) :$ 
\begin{equation}
k\left( \theta ,t\right) =-\eta \left( \theta ,t\right) \frac{\partial \psi
\left( \theta ,t\right) }{\partial \theta }=\eta \left( \theta ,t\right) 
\frac{\partial }{\partial \theta }\frac{1}{2\pi }\int\limits_{0}^{2\pi
}d\theta ^{\prime }\coth \frac{\theta ^{\prime }-\theta }{2}\ln \eta \left(
\theta ^{\prime },t\right) .  \label{f29}
\end{equation}
One can also notice \cite{glb,ben,gl2}\ that 
\begin{equation}
\upsilon _{n}(\theta ,t)=\eta \left( \theta ,t\right) \left( 1-d_{0}\frac{%
\partial }{\partial \theta }%
%TCIMACRO{\func{Im}}%
%BeginExpansion
\mathop{\rm Im}%
%EndExpansion
\left. \left[ S\left[ k\left( w,t\right) \right] \right] \right|
_{w=e^{i\theta }}\right)  \label{f30}
\end{equation}

Equations (\ref{f28})-(\ref{f30}) represent the system of
intrgrodifferential equations that fully describes the evolution of the
interface.

To solve the derived singular integral equations with Hilbert kernel, the
useful approach is developed, based on interpolation of the expressions of
integration in singular integrals by trigonometric polynomial. For example,
for each number $N=2n\;$of the nodes we can use$\ $the formulae \cite{pns}

\begin{equation}
\widetilde{f}_{n}(\theta )=\frac{1}{2\pi }\sum_{j=0}^{2n-1}\widetilde{f}%
(\theta _{j})\sin [n(\theta -\theta _{j}))\cot \frac{\theta -\theta _{j}}{2}
\label{f31}
\end{equation}
\begin{equation}
\theta _{j}=\frac{\pi j}{n},\qquad j=0,1,...,2n-1  \label{f32}
\end{equation}
which lead to the next integral representation:

\begin{equation}
\frac{1}{2\pi }\int\limits_{0}^{2\pi }d\theta ^{\prime }\coth \frac{\theta
^{\prime }-\theta }{2}\widetilde{f}(u_{j})=\frac{1}{2n}\sum_{j=0}^{2n-1}%
\widetilde{f}(\theta _{j})\cot \frac{\theta _{j}-\theta _{m}^{0}}{2}
\label{f33}
\end{equation}
where 
\begin{equation}
u_{m}^{0}=\frac{2\pi m+1}{2\pi },\qquad m=0,1,...,2n-1  \label{f34}
\end{equation}

The system of nonlinear integrodifferential equations was investigated
numerically. The basis of the numeric simulation was the approximation of
unknown quantities of functions by above mentioned trigonometrical
polynomial. Expansions of this type make it possible to use well-known
quadrature formulae to evaluate singular Integrals . As a result, the system
of equations (\ref{f28})$,$(\ref{f29}) and (\ref{f30}) reduces to a system
of ordinary non-linear equations with full Jacoblan.

It is possible to reproduce geometrically defined interfaces in parametric
form by using the equations:

\begin{eqnarray}
dy &=&\frac{d\theta }{\eta (\theta ,t)}\sin \left( \theta +\psi (\theta )+%
\frac{\pi }{2}\right)   \label{f35} \\
dx &=&\frac{d\theta }{\eta (\theta ,t)}\cos \left( \theta +\psi (\theta )+%
\frac{\pi }{2}\right)   \nonumber
\end{eqnarray}

In numerical experiment the evolution of the form of an interface of phases
in an assotiation from an aspect of initial perturbations $\delta \eta
|_{t=0}$ of a surface energy $d_{0}$ was investigated.

Figures (\ref{fig2}) and (\ref{fig3}) illustrate the characteristic form of
non-uniform distributions of the $\eta ,$ $\psi $ and the corresponding
interfaces for different initial perturbations. The reader should observe
that relatively smooth variations of the interface correspond to quite
marked variations in the field $\eta $ and hence also in the angle $\psi .$
Thus, relatively small errors in determining do not cause marked changes in
the interface geometry. In addition, as is evident from the figures,
relatively even long sections of the Interface contract when mapped into the
interval $\left[ 0,2\pi \right] $, whereas pronounced changes in the contour 
$\Gamma $, conversely, are stretched out. In this sense such mappings of
onto are ''adaptive'' with respect to information about the detailed
geometic structure of the interface.

\begin{figure}[tbp]
\caption{Non-uniform distributions of the $\protect\eta \left( \protect\theta
,t\right) ,$ $\protect\psi \left( \protect\theta ,t\right) $ and the
corresponding interfaces for the following initial perturbation: $\protect%
\eta \left( \protect\theta ,t_{0}=0\right) =0.1-0.01\cos (4\protect\theta );$
$d_{0}=0.1$; $t_{1}=500;$ $t_{2}=1000;$ $t_{3}=3000$; $t_{4}=6000.$ Bold
interfaces on the figure correspond to mentioned above instances of time.}
\label{fig2}
\end{figure}

\begin{figure}[tbp]
\caption{Non-uniform distributions of the $\protect\eta \left( \protect\theta
,t\right) ,$ $\protect\psi \left( \protect\theta ,t\right) $ and the
corresponding interfaces for the following initial perturbation: $\protect%
\eta \left( \protect\theta ,t_{0}=0\right) =0.1-0.01\cos (6\protect\theta );$
$d_{0}=0.1$; $t_{1}=500;$ $t_{2}=1000;$ $t_{3}=3000$; $t_{4}=6000.$ Bold
interfaces on the figure correspond to mentioned above instances of time.}
\label{fig3}
\end{figure}

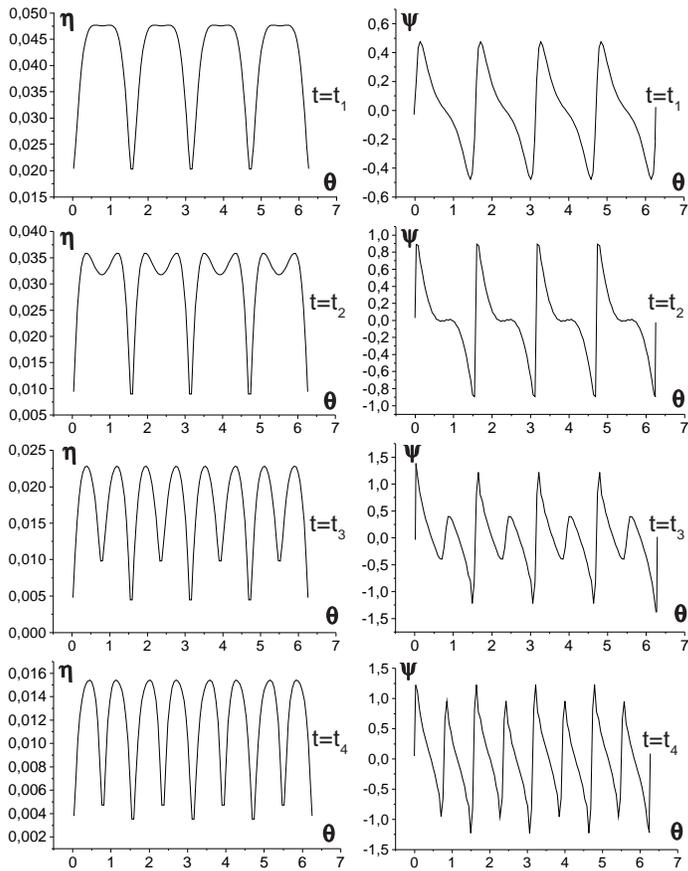
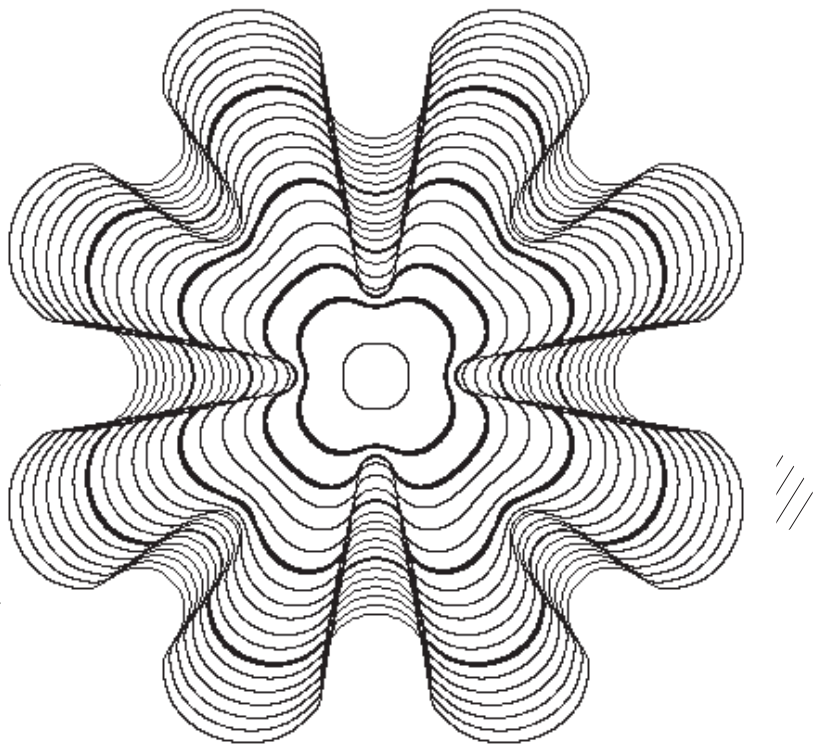
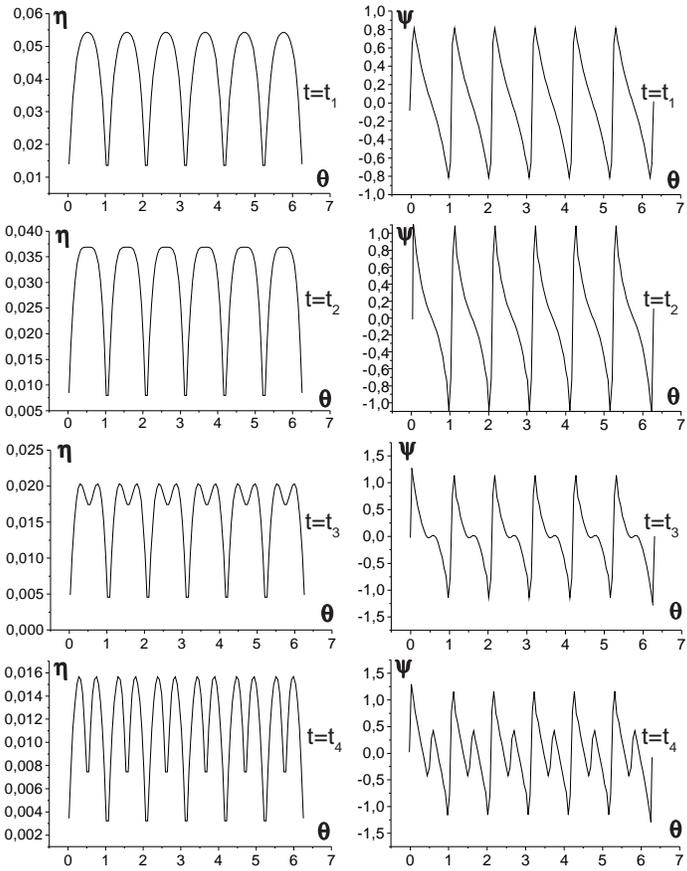
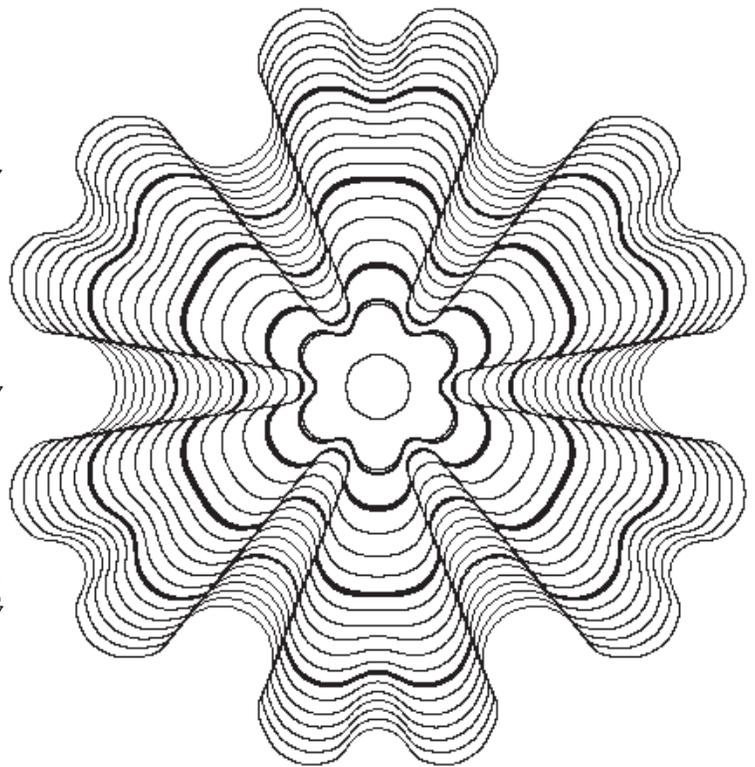
\end{document}